# Evolution of Cooperation, Differentiation, Complexity, and Diversity in an Iterated Three-person Game

Eizo Akiyama*and Kunihiko Kaneko†

## Abstract

A non-zero-sum 3-person coalition game is presented, to study the evolution of complexity and diversity in cooperation, where the population dynamics of players with strategies is given according to their scores in the iterated game and mutations. Two types of differentiations emerge initially; biased one to classes and temporal one to change their roles for coalition. Rules to change the hands are self-organized in a society through evolution. The co-evolution of diversity and complexity of strategies and interactions (or communications) are found at later stages of the simulation. Relevance of our results to the biological society is briefly discussed.

## 1 Introduction

In a society with inter-acting agents, emergence of cooperation is commonly observed, while the diversity and complexity there are increased through class differentiation or temporal changes of roles. In the present paper we discuss the mechanism of such evolution by adopting an iterated three-person game.

The evolution of cooperative behaviors observed among selfish individuals has been a topic of debates over decades, especially among evolutionary biologists. There are two hypotheses, the genetical kinship theory and the reciprocity theory, which explain the origin of such cooperation. The kinship theory[2] gives satisfactory explanations about altruistic behaviour in honeybees, ant workers and so on. On the other hand, individuals without blood relationship one another have to recognize and attain the cooperation in the reciprocity theory, where Iterated Prisoner's Dilemma (IPD) model is most popularly studied. In the Pisoner's Dilemma (PD) game, two players either cooperate or defect, with the score in Table1. Computer tournaments of IPD programs were organized by Axelrod, where each player has a strategy depending on the history of hands [1]. The most successful strategy therein was well-known Tit-For-Tat (TFT), that cooperates on the first move and then plays whatever the other player chose on the previous move. When the evolution of strategy is included, the cooperation prevails in society, without any explicit indication, through the success of the TFT algorithm. Thus the emergent society is cooperative, in which the strategies therein are very simple and basically uniform by players.

In the nature and in our society, the form of cooperation is not necessarily such simple. Actions are not always uniform in time or by players. In the cooperation with temporal changes of actions, which we call temporal differentiation here, players change their roles through some rules. Such differentiation is seen in the following examples:

- In a shoal of fish such as sardines and herrings, the risk eaten by larger fish are higher in perimeters. They frequently change their position and direction, and share the risk.
- In a recent model of cell differentiation with competition for nourishment among cells [4], they actively take the foods or rest in turn, to form a kind of time sharing system (while the biased differentiation is observed at later stage). Such temporal differentiation is also seen in experiments with E.Coli[5].

On the other hand, cooperation only among a part of the members in a group is seen for example in the following cases:

- In a group of birds, only a certain sub-group makes alarm calls to tell other members the existence of predators.
- A small group of fish takes the risk of inspecting potential predators. (e.g.[7])

To study such forms of cooperations with differentiation of roles, a 2-person game is not adequate. For this we introduce and study a simple non-zero-sum 3-person game model here.

### Complexity and diversity in strategy and communication

Another drawback in the IPD model is the lack of complexity and diversity. For the formation of cooperation, there must be some kind of communications. In the IPD model players communicate only through the information of the history of hands, obtained by the repetition of games. In the evolution of IPD model, however, the final society is very simple with the actions Cooperate only ( in some special cases Defect only), and the society is dominated only by the TFT-like strategies. Thus the model cannot explain the diversity and the complexity in our world, where various forms of communications and strategies coexist, ranging from simple to sophisticated ones.

*Department of Pure and Applied Sciences, University of Tokyo, 3-8-1 Komaba, Tokyo 153, Japan, E-mail: akiyama@complex.c.u-tokyo.ac.jp

†E-mail: kaneko@complex.c.u-tokyo.ac.jp

Table 1: the pay-off matrix for Prisoner's Dilemma : In each element, $(S_1, S_2)$ corresponds to the score of player 1 and player 2, respectively.

| | | player 2 | |
|---|---|---|---|
| | | **C** | **D** |
| player 1 | **C**ooperate | 3, 3 | 0, 5 |
| | **D**efect | 5, 0 | 1, 1 |

One possible way to get rid of the drawback may be the inclusion of noise, as player's errors of actions, as has been studied by Lindgren[6]. Through the evolution, the memories of previous hands are increased in the strategy, after alternations of dominant strategies. In this model, however, the action is still "Cooperate" only (except for some intervals to get rid of the noise effect), at later generations. The strategies are still dominated by long-term versions of the TFT. Thus the noise effect is not adequate to account for the complexity and diversity.

Of course, a straightforward way to introduce the complexity is by combinatorics, and is to include a variety of hands in the game, like the chess. We do not take this direction however, since we are interested in the origin of diversity and complexity solely through the inter-actions of players, *without* implementing it in a game initially. Thus the use of a three-person game is again requested as a possible simplest model at the next step.

### N-person game

There is a qualitative difference between 2-person and N-person games. (N $\geq$ 3) It is mainly due to the possibility of more than two coalitions. In an N-person game, there are variety of partitions of players into sub-groups forming coalitions. To form coalition, some communications are necessary which may take complex and diverse forms, as are made possible by temporal changes of roles in the coalition.

In the present paper we study the simplest n-person game, a deterministic 3-person, and non-zero-sum game with two hands, focusing especially on the structure of the coalition. The evolution of artificial ecology of species with different strategies is studied through repeated games by players. The main topics to be discussed are

- emergent forms of cooperations
- the evolution of algorithms and communications
- the dynamics of diversification and complexification
- the nature of the society evolved.

Indeed our simulation shows class differentiations between exploiting and exploited players at the initial stage, and then the temporal differentiation of roles to attain the cooperativity. At later stages the co-evolution between the complexity and diversity is found for communications and strategies.

# 2 Modeling

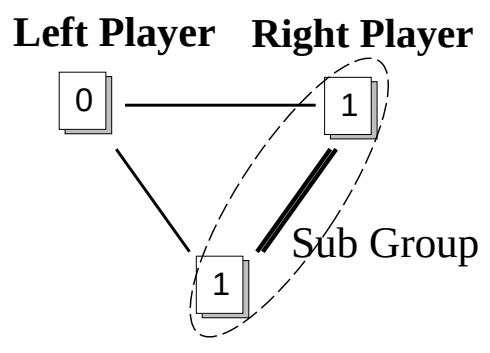

Figure 1: sub-group with the right player

The rule of our three-person game is as follows;

1. Each player must hand in either card 0 or 1.
2. If two players hand in the same cards, they are regarded as forming a sub-group, and gain score (3 points). A player excluded from the sub-group cannot gain any score. If all the three players hand in the same cards, they cannot get any score, either.

The pay-off matrix is given in Table 2. In the table, we distinguish right and left players, assuming that the three players are located in a circle so that each player has its right and left player. Of course the rule of our game keeps the right/left symmetry. However, each player is assumed to be able to distinguish the right and left players, which is essential to the choice of its strategy, as will be seen later.

In each round of the game, a player hands in the cards repeatedly in succession until a given maximum round number is reached. Such iterated actions of three players as a whole will be called simply as 'interaction'. The cards to be handed in by each of the players are decided according to their strategy, referring to the history of the states, defined by the hands of the three players as in Table2. The coding of the strategy algorithm is given by defining an octonary tree structure according to the history of the states, as in the binary tree coding by Ikegami[3].

Table 2: pay-off matrix of our 3-person coalition game : The number '0' or '1' in column 2, 3, or 4 represents respectively the card that the left player, the right player, or you have handed in. According to the hands of the three players, there are 8 states, which are defined through the binary representation of their hands, as is given. If and only if your state is between 2 and 5, you are in a sub-group and can get 3 points.

| state | Left | Right | You | point |
|---|---|---|---|---|
| 0 | 0 | 0 | 0 | 0 |
| 1 | 0 | 0 | 1 | 0 |
| 2 | 0 | 1 | 0 | 3 |
| 3 | 0 | 1 | 1 | 3 |
| 4 | 1 | 0 | 0 | 3 |
| 5 | 1 | 0 | 1 | 3 |
| 6 | 1 | 1 | 0 | 0 |
| 7 | 1 | 1 | 1 | 0 |

The memory-length, that is the number of prior rounds to be referred for the strategy, is provided for each algorithm, which is finite within a given fixed range. Thus, the next card to be handed in is decided according to the finite length history of states. Also, the information of the first card is given in each player's algorithm. Figure2 is an example of the game play, where the player 1, 2 and 3 are located in an anti-clockwise order, and the state is decided according to Table2.

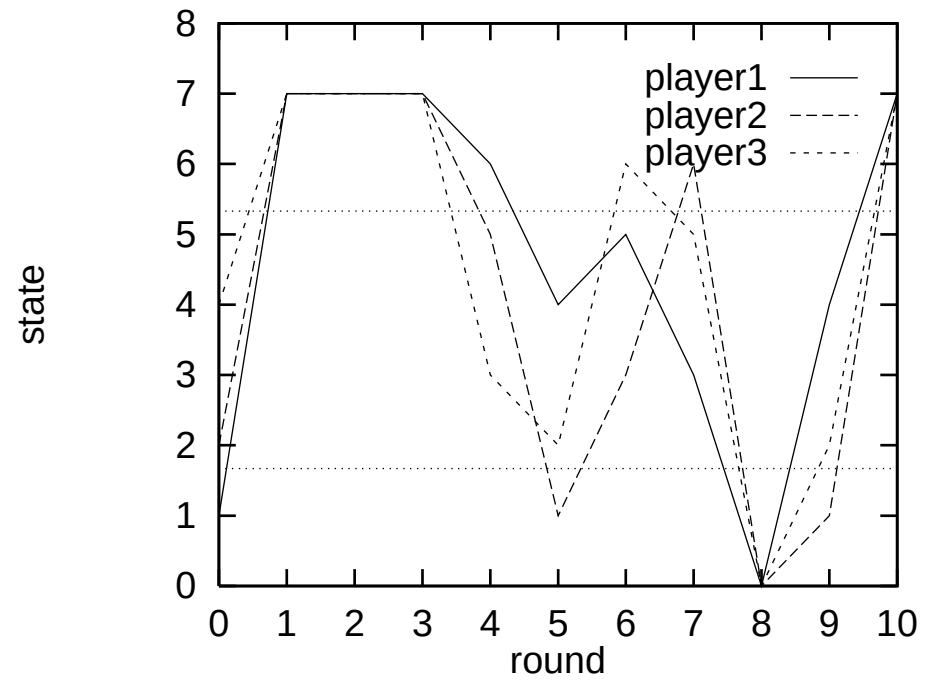

Figure 2: An example of the game play : The horizontal axis shows rounds, while the the vertical axis shows the states of the players. Dotted lines are drawn near states 2 and 5. Players whose state is between these dotted lines can get 3 points.

By taking an ensemble of players and regarding the players with the same strategy as the same species, we study the population dynamics of each species. The population dynamics is defined as follows: In each step, a player makes the 3-person game with all possible pairs of the other players, including those from its own species. By summing up the points of the game, a player's, accordingly the species', score is given. The fraction of the population $x_i(t)$ for a species $i$ is updated, following its average score $s_i$ subtracted by the average points of all the players $\overline{s}$ [1]:

$$x_i(t+1) - x_i(t) = d\,(s_i - \overline{s})\,x_i(t) \qquad (1)$$

where d is a growth constant. After all the fractions $x_i(t)$ are updated, they are normalized to make the population size 1.0.

When the population is updated to the next generations, a single point mutation of the algorithm occurs with a given fixed ratio (0.1 in later examples). Here the mutation adds or removes one branch at every node in the tree of the algorithm.

As noted previously, both the game and the algorithm are deterministic. Thus all the three players of the same species hand in the same cards for each round. Since the state and the memory-length are finite, the change of the state must finally fall in a periodic cycle through the iteration of the game. The rounds showing the cyclic change of states will be called a periodic part, while those before the periodic part will be called a transient part.

### Differentiation of roles

In order to gain some points, the cooperation of 3 players is necessary, where one of the other two players gives in and plays the role of an outsider from the sub-group, or that of 2 players making the remaining player the odd man out. In either case, three players must split into two and one, to gain points.

As will be observed from the result of the simulation, there are two ways of differentiations.

**Class differentiation** A biased differentiation. The roles are fixed by players, and a particular player loses on the average and is exploited by others. ( Figure3-(a) )

**Temporal differentiation** The roles of players to form coalitions change with time. ( Figure3-(b) )

If one of the players is out of the coalition in turn, each player gets 2 points on the average, and the full and equal cooperation is attained. If the interaction is not far from this ideal situation, we call it **cooperative interaction**, where the average scores are high ( close to 2), and their difference by players is small. The interaction by the temporal differentiation provides a typical example.

[1] If the score of a species is below the average $\overline{s}$ and its population goes down below a given value (KillLimit), it is assumed to be extinct, and the species is eliminated.

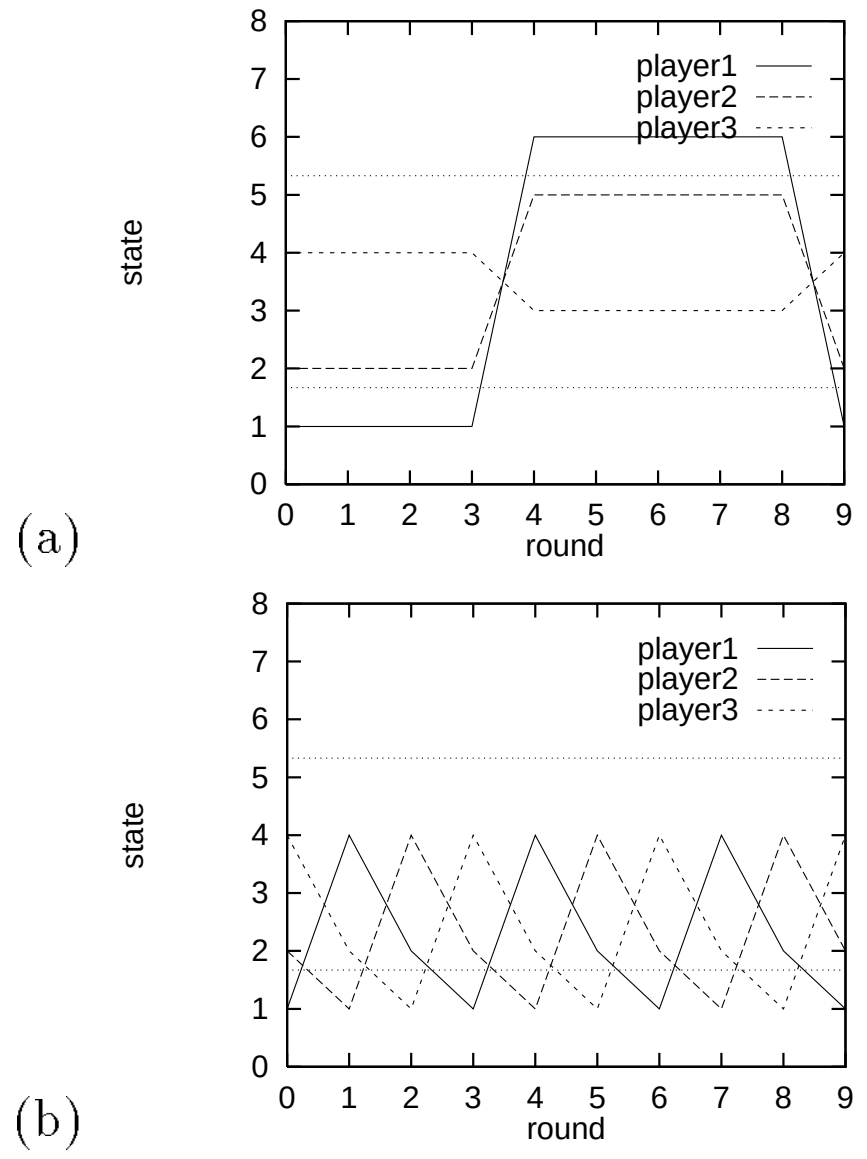

Figure 3: (a) Example of class differentiation (b) Example of Temporary differentiation

It may be useful to note the important difference between our game and the PD, besides the number of players. In the PD, the two hands, C(ooperate) or D(efect) have their specific meaning, and the game is asymmetric between C and D. Thus the evolved strategy as well as the action should strongly depend on C or D. In our 3-person game, the hands 0 and 1 themselves are symmetric and have no specific meanings. Information to make some kind of communication and form a sub-group is given in the time series of the hands. As will be seen, societies of various types of periodic hands such as the period-3 of 001 or period-5 of 00101 are formed through the evolution.

# 3 Simulation results

We have carried out simulations of the 3-person game, setting the maximum round number to 1000, and the maximum memory length to 4. The simulation starts with 6 species whose algorithms are given by the tree made randomly with the memory length = 1.

First we present a rough sketch of the evolution of our model while detailed accounts will be given later. Through several simulations, we reach the following scenario of the evolutionary process to the complex and cooperative society :

1. A new species arising from mutations leads to class differentiation, which lowers the score of the old species and its population. Thus the society is tended to be dominated by the new species.

2. This dominance is broken by the emergence of cooperative interactions, supported by periodic temporal differentiation. The ratio of cooperative interactions increases with the evolution.

3. The temporal differentiation of periodic changes of hands with the $3n$ period ($n = 1, 2 \cdots$) dominates the society. The whole species therein shows the identical patterns of the hands at the periodic part, while the diversification occurs only in the transient part.

4. Some mutants which also change the periodic part increase their population, and dominant periods in the society are changed. After having experienced alternations of some dominant periods, the society starts to allow for the coexistence of various periods. With this increase of diversity, the interaction and strategy increase the complexity, through the appearance of longer periods.

All the simulations support the above evolutionary process, although there are subtle differences by simulations in the period of the cyclic change of hands and the order of societies realized.

## 3.1 Class differentiation and the emergence of cooperation

### Class differentiation

In our 3-person game, some kind of rules, such as the periodic exclusion from a sub-group, must be formed by the players, to gain points by the game. Such rule, however, is not easily formed. Except for some special initial conditions which allow for such cooperation by chance, some players are ignorant of the rule of cooperation, and are exploited by others. This leads to the class differentiation. Since those exploit others get higher scores, the exploitation is increased through the evolution. With generations, new species with a longer memory length appears which adopt a more complicated rule to exploit others. Thus the class differentiation with a more complex strategy emerges successively.

The simplest example of class differentiation is shown in Figure4-(a), where two players of the same species handing in the card 0 exclude the remaining player from the sub-group. (Note again that the player of an even state hands in card 0, while that of the state hands in card 1. See Table2.) In this simple case, the excluded player could have escaped this exploitation, if it adopted a simple 2-memory-length strategy like "if excluded twice by the same cards, change the hand". Indeed this type of mutation occurs at a later stage, while there appears a more complicated form of the exploitation as in Figure4-(b)(c) by using a longer memory length.

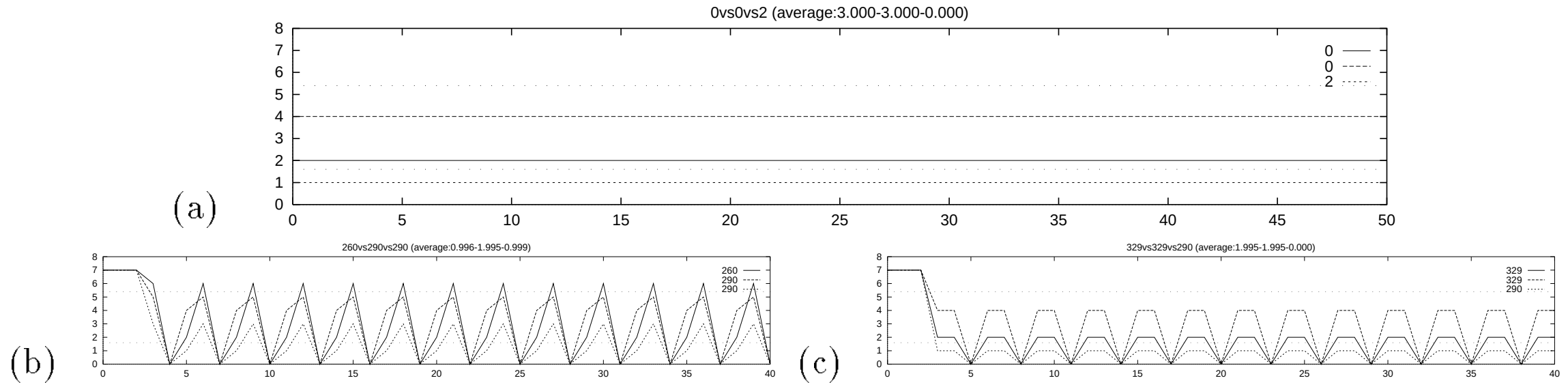

Figure 4: Examples of Class Differentiation : In (a), the three players are ID0, ID0, ID1—two persons from species ID0 and one from species ID2—, get the average scores 3.000, 3.000, 0.000 respectively. (b) and (c) are examples from later generations.

For example, in (c), the player changes cards when excluded twice, but still it is exploited with the rate 2/3. The excluded player again can escape from the exploitation by having a longer memory length of the strategy (4 in this case). Thus the complexity is increased within the class differentiation, where a species with a complex strategy (with a longer memory length) dominates over a long time.

### Emergence of Cooperation and its evolution

In the class differentiation, the dominant species increases their population by exploiting other species. Thus, when the species occupies most populations of the society, it cannot get scores any more. If there appears a new species that is not exploited by the dominant one and cooperates with each other (Figure5-(a)), its relative population is increased. Thus the society of class differentiation collapses, after which the cooperation expands in the society. An example is given in Figure 5, where in (c) at a later generation, the players get points by the cooperation with periodic differentiation of roles.

## 3.2 Temporal differentiation

### stable and uniform society with temporal differentiation of period-3n

After the emergence of cooperative interactions by the temporal differentiation, the society with the period-3n is gradually formed, where the players equally exchange the role of the excluded. Furthermore, any set of three players from different species perform the same period-3n changes of hands.

An example is given in Figure6-(a)(b), where society with period-6 interaction emerges, and continues stably over many generations. Here, the subgroup with the hand 1 is formed, and each of the three players is excluded twice per 6 steps, by showing the hand 0. As shown in this example, all the three players get the equal score in the period-3n society.

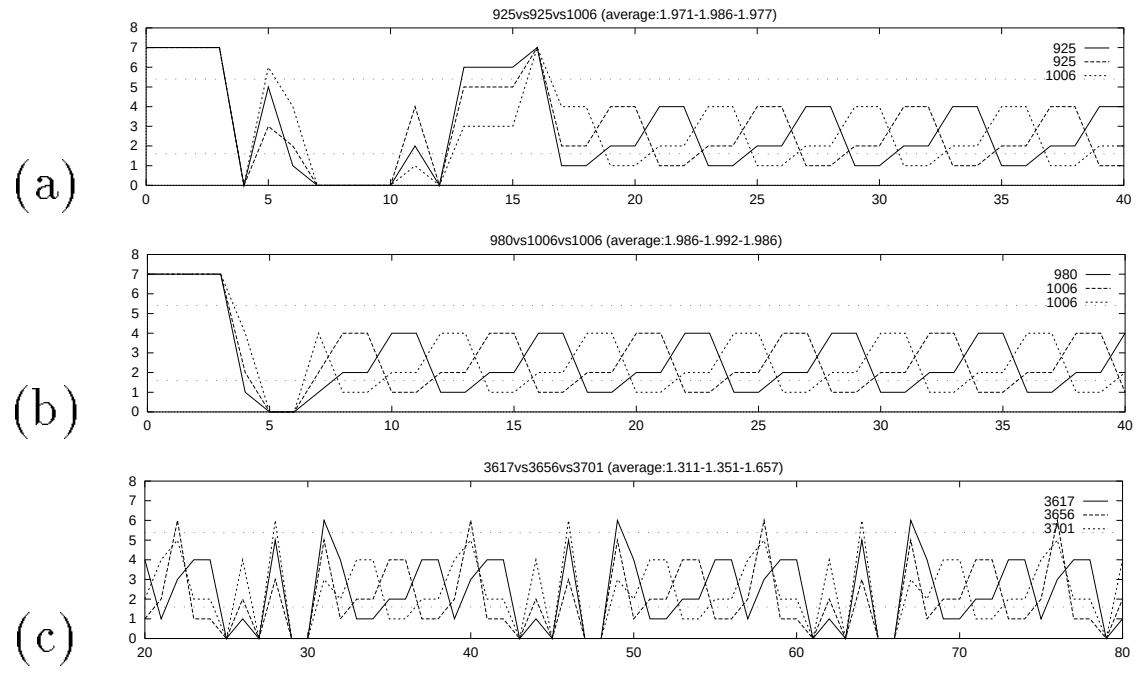

Figure 6: Temporal differentiation in a uniform period-6 society : (a) and (b) are representatives of inter-actions in a society where all interactions are period-6. (c) is a sample of the pseudo period-6 interaction, which will destroy the period-6 society.

### The diversification of the transient parts

Since each player in the period-3n society gets the highest possible score among "equal-score" societies, it is rather difficult for a new species to exceed the predecessors by adopting a different type of periodic patterns. The easiest and commonly observed strategy of a mutant at this stage is to preserve the periodic part and change the transient part, during the equal cooperation is not attained.

It should be noted that the transient part is essential to shift the phases of the period-3n oscillation by players, since they should change the hands out of phase each other, to form the cooperation. There can be a variety of choices for the transient part. Indeed in our simulation, new species with modified transient parts appear successively.

### New species with interactions indiscernible by old species—the end of the uniform period-3n society

As the species of period-3n society gain points efficiently, a new species with modified periodic parts have to exploit the old period-3n species, to expand

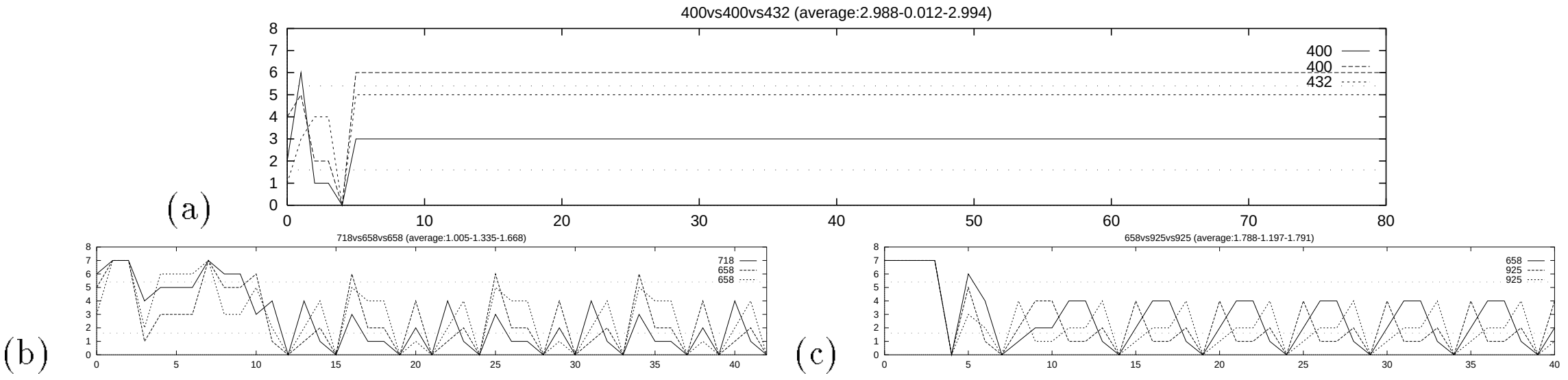

Figure 5: The emergence of cooperation and its evolution : (a) A (right) player from species ID400 hands in card 0, while the left player from ID400 and one from ID432 hand in card 1. Thus, both species can get gains. This is an example of imperfect cooperation where two species gain unequal average score. In (a) to (c), the increasing element of temporal differentiation, by which participating species get more even score, in each interaction can be observed,

its population. Since such mutation is not easy ( and indeed the period-6 society last over many generations), it emerges only after a long enough period. First, new species appears which shows the same period-3n interactions fundamentally, but shifting the phase to a degree undetectable by the old species. We will call such type of interactions as 'the pseudo period-3n interactions', in which the periods are longer than 3n but some fractures of the original perio-3n interaction is included. In fact, as is shown in Figure6-(c), the new species lowers the old species' points while retaining to a degree its own points by performing 'the pseudo period-6 interactions'. The original period-6 species cannot prevent this attack.

**The transition of societies with periodic temporal differentiation**

The evolution that undermines periodic parts itself continues, even after the termination of the period-3n society. At this stage, a variety of periodic interactions appear successively.

## 3.3 Diversification and complication

### Evolution to diversification

So far the society is composed mainly of one type of inter-actions (with the same period). The diverse inter-actions are unstable and observed only in the transition between stable societies. At this late stage, however, the society consists of several different inter-actions with different periods, and remains stable.

Such society appears first in our simulation as the coexistence with period-3 and period-6 interactions (See Figure8-(a)), which is born out of the period-3 society. After some generations, a variety of inter-action co-exists as in Figure8-(b)(c),all of which are examples of inter-actions chosen from three players in the same society.

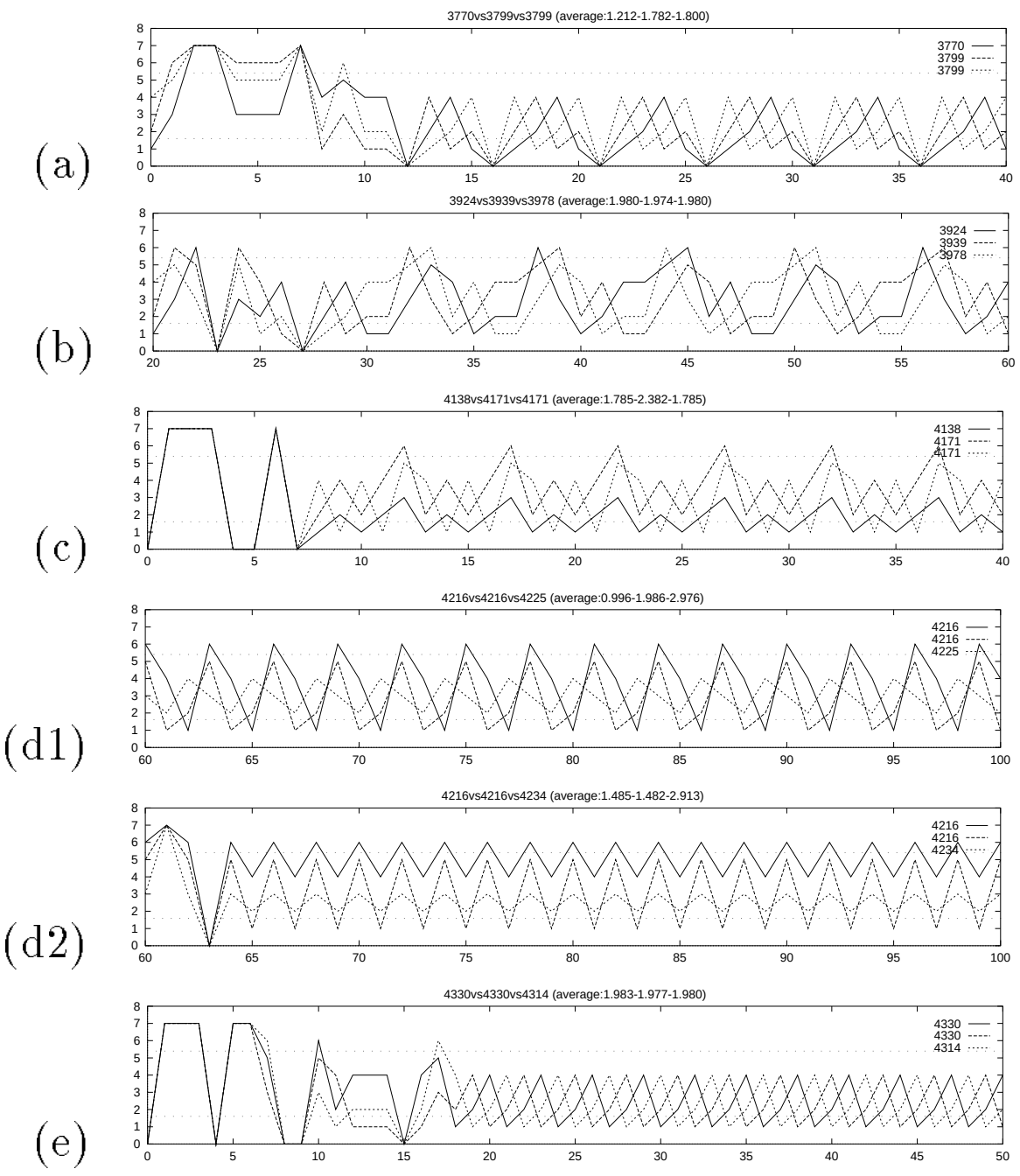

Figure 7: The successive change of periodic societies with generation : (a) is a sample of interactions from period-5 society, (b) from period-18, (c) from another type of period-5, (d) from a diversified society lasting only for a short span, (e) from period-3.

### Evolution to complexity by breaking the phase of hands

The strategy to break the phase of the oscillation, already seen in the pseudo-3n inter-action, is again seen here. In contrast with the pseudo-3n case, however, more complex inter-actions emerge successively by breaking the phases more frequently. For example, in Figure9-(b)(c) the periods for the cyclic hands are about 100. The dynamics here is rather irregular, and looks rather unpredictable. We note that even in this society, some of the inter-actions remain very simple, such as the period-6 one as in Figure9(a). Complex inter-actions exist within the diversity of species, while the diversity is supported by the complexity in strategies.

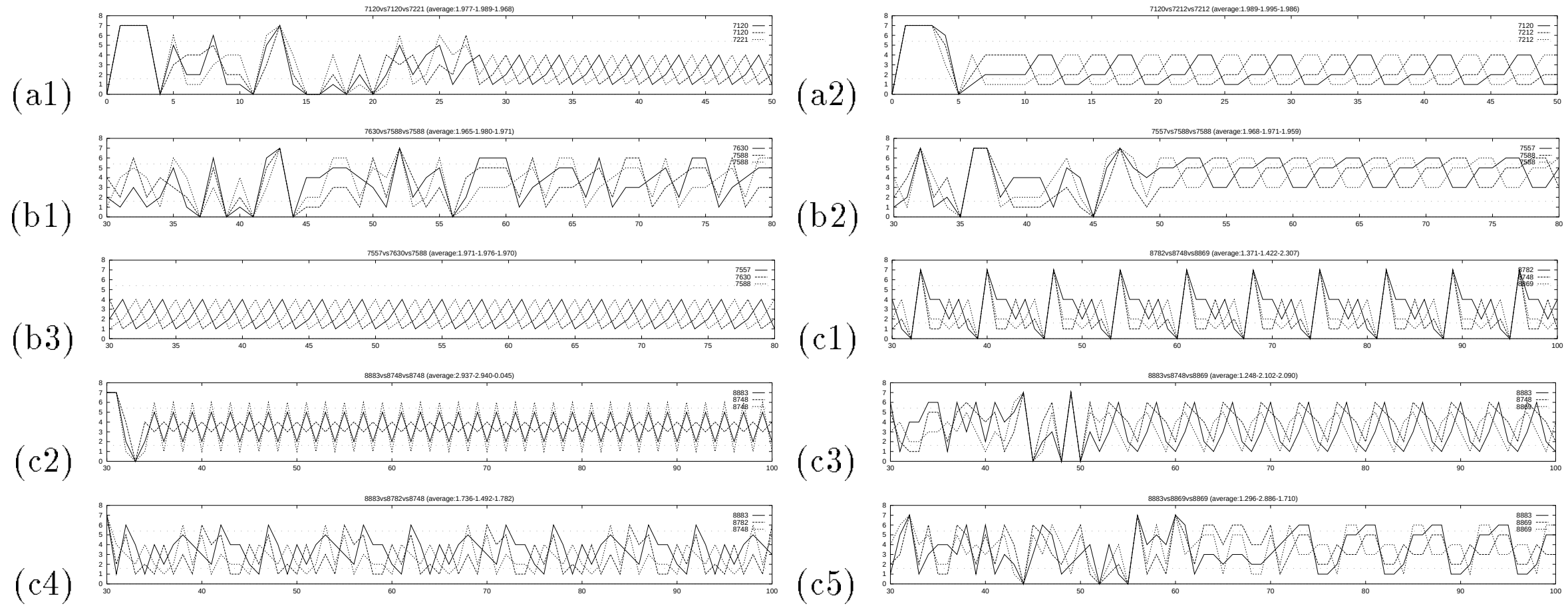

Figure 8: The evolution to diversification from (a), (b), to (c) in time : (a1–2) co-existence of period-3 and 6 inter-actions in a society. (b1–2) that of period-3, 6, 15. (c1–5) a variety of inter-actions seen in a society.

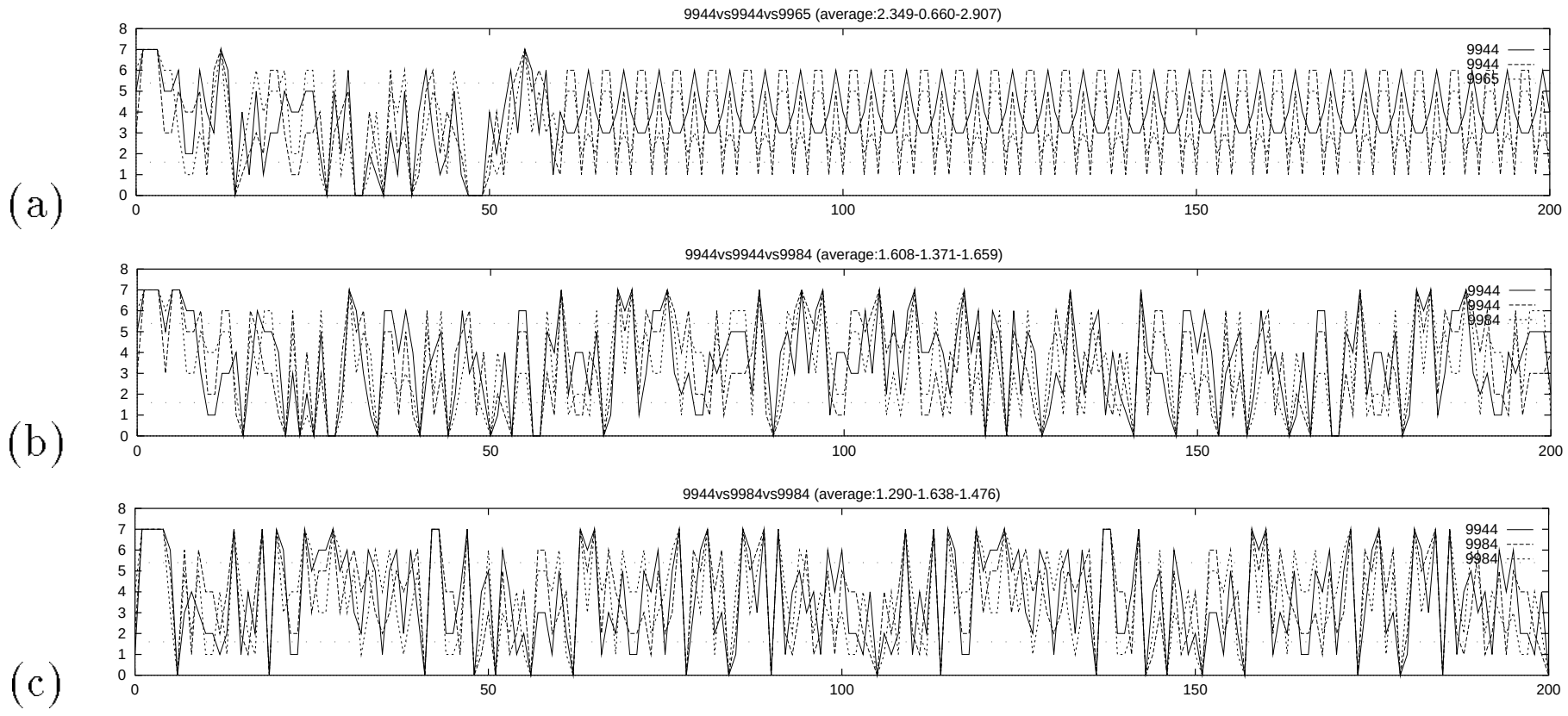

Figure 9: Diversification and complication : (a)–(c) are representatives of inter-actions in a society. (a) is rather simple, while others are quite complex with periods about 100.

# 4 Discussion

In our deterministic three-person coalition game, we have found the evolutionary process starting from class and temporal differentiations of roles, and reaching the diversification of society and the complexification of interactions. We note such evolutionary process has not been found in simulations of 2-person games such as the IPD model.

### The emergence of cooperation due to temporal differentiation of roles

When resources are in scarcity, or when some player must bear a dangerous role, some player must suffer loss of profits. In this case, we have found two types of differentiations to resolve such situation, class and temporal differentiations.

In the class differentiation, a caste society is formed where only certain parties continuously suffer some loss. In such society, the lower class is finally extinct, and a new exploited class should be formed. This process must be repeated forever to preserve a class society, which is rather improbable. Indeed, in our simulation, such society lasts only for some time, and is typically unstable. The society finds another solution, the dynamic change of roles, which is cooperative and lasts as a stable state over many generations.

In our 3-person coalition game, card 1 and 0 themselves have no specific meanings, and are symmetric. Some logic to break the symmetry and to assign meanings to the dynamics of hands is self-organized by forming rules of societies through the evolution. Here the formation of rules is partly trig-gered by the ability of players to distinguish the (right/left) position. We have also made several simulations without this ability ( in other words, using the algorithm depending only on the numer of 0 and 1 by the other two players, besides its own hand), where we have found only the class differentiation, but not the temporal one. Thus the temporal differentiation seems to be formed through the ability of the location discrimination, for example, by the rule that 'each player should give in if its right player gave in in the previous round (leading to a clockwise period-3 society).

### Diversification of the society

Societies with the temporal differentiation of various periods coexist, at a later stage in our simulation. Clockwise and anti-clockwise, and cooperative and exploiting interactions coexist with different periods such as 3, 6, and 15. This diversity is possibly provided by the existence of many "metastable" solutions without an absolute advantage of any group, in our 3-person game, in which no a priori advantage of hands (1 or 0) is implemented. Indeed such diversification is not found in the simulation of IPD model (with/without noise), where a stable society shows cooperative actions only, with one (or few) dominant strategy.

In reality, as in human society, there are diverse forms of cooperations, while not all the individuals participate in the cooperation itself. For example, not all birds make the alarm calls discussed in section 1. Of course, studies of an n-person game required for the alarm call of birds, but the observed diversity in our 3-person game gives a useful suggestion for future studies.

### Evolution of Complexity

With the evolution, more complex interaction with longer periods has been observed. The increase of memory length, so far, has been observed in the IPD model with a noise, although the action itself is not complex there ( always 'cooperate' unless noise is added). In this respect, the emergence of longer memory in our deterministic 3-person game may suggest that the third player may play a kind of role of "noise", in the course of the evolution.

However, the complexity in our game does not only lie in the long memory length but also in the inter-action itself ( or actions). Indeed there are two other essential mechanisms for the complexity: One is the competition between exploiting and avoiding being exploited. Simple rules for the coalition are easily detected by others, and may be exploited by a more complex one. Thus there appears a pressure for developing a complex strategy and interaction. This mechanism of the evolution is common with that observed in the imitation game [8], and reminds us of the evolution of (secret) communication codes: Those decoded only within the same group may be generated through complexity.

Another mechanism is related with the diversification. In a diverse society a player has to cope with a variety of interactions. A simple strategy cannot afford such diverse responses, including for example, denial of coalition with some players, avoiding the coalition of 3 players by temporal differentiation, and a coalition of 2 players by class differentiation. Thus diversity enhances (temporal) complexity of interactions, while the diversity itself is supported by the complexity of strategies, since the diversity of interactions is limited if their periods are short. Thus the diversity and complexity of interactions and strategies *co-evolve* in our simulation, which seem to be seen in (real) ecological systems and in human society.

## Acknowledgments

The authors would like to thank T.Ikegami, S.Sasa, and T.Yamamoto for useful discussions.